%% file: rayleigh_honeycomb_lattice_mtt_arxiv2.tex
\documentclass[11pt]{article}

\usepackage[top=1.0in,right=1.0in,left=1.0in,bottom=1.75in]{geometry}
\usepackage{amssymb,amsbsy}
\usepackage{amsmath}
\usepackage{amsthm}
\usepackage{graphics,graphicx}
\usepackage[T1]{fontenc}
\usepackage{color}
\usepackage{subfigure}
\usepackage[affil-it,auth-sc]{authblk}
\usepackage{stmaryrd}

\usepackage[]{jabbrv}

\usepackage[numbers,sort&compress]{natbib}

\usepackage{setspace}

\usepackage{soul}
\definecolor{lightyellow}{rgb}{1.0, 1.0, 0.88}
\definecolor{lightmauve}{rgb}{0.86, 0.82, 1.0}
\sethlcolor{lightmauve}

\usepackage{titlesec}

\titleformat{\section}[runin]
{\normalfont\bfseries}
{\thesection.}{0.5em}{}

\titleformat{\subsection}[runin]
{\normalfont\bfseries}
{\thesubsection.}{0.5em}{}

\input{definitions_ijss}

\graphicspath{{./}{./figures/}}

\title{Floquet-Bloch waves in periodic networks of the Rayleigh beams: \\ honeycomb systems, dispersion degeneracies and \\ structured interfaces}

\author[1]{L. Cabras}
\author[2]{A.B. Movchan\footnote{Corresponding author: e-mail: abm@liv.ac.uk}}
\author[1]{A. Piccolroaz}

\affil[1]{Dipartimento di Ingegneria Meccanica e Strutturale, Universit\`a di Trento, Italy}
\affil[2]{Department of Mathematical Sciences, University of Liverpool, U.K.}

\numberwithin{equation}{section}

\date{}

\begin{document}

\maketitle

\centerline{\em In honour of Professor N.F. Morozov, on the occasion of his 85th Birthday}

\vspace{.2in}

\begin{abstract}
\noindent
The paper addresses novel dispersion properties of elastic flexural waves in periodic structures which possess rotational inertia. The structure is represented as a lattice, whose elementary links are formally defined as the Rayleigh beams. Although in the quasi-static regime such beams respond similarly to the classical Euler-Bernoulli beams, as the frequency increases the dispersion of flexural waves posses new interesting features.
For a doubly periodic lattice, we give a special attention to degeneracies associated with so-called Dirac cones on the dispersion surfaces as well as directional anisotropy. Comparative analysis for Floquet-Bloch waves in periodic flexural lattices of different geometries is presented and accompanied by numerical simulations. 
\end{abstract}




%

\section{Introduction.}
\label{sec01}


Dynamics of elastic lattices is of exceptional importance in a wide range of applications in problems of structural mechanics. 
The pioneering work of N.F. Morozov and his co-authors has addressed scale effects and dynamic  trapping for cracks propagating in transient regimes through structured media, as well as formation of nanocracks in crystalline solids \cite{MPU1, MP, MPU2, Morozov_Petrov,Morozov_Petrov1}. An important concept of the dynamic fracture toughness at a stage of the crack initiation has been fully investigated, and an ``incubation time'' principle has been introduced by N.F. Morozov and his co-authors to describe time-dependent fracture in structured materials for the general types of loading. 





A significant impact has been made by the investigations of N.F. Morozov {\em et al.}\ \cite{MT1,MT2,MT3,MT4} on the dynamics and stability of elastic rods subjected to transient longitudinal loads.   A new insight has been presented by Morozov and Tovstik, who had addressed the influence of longitudinal loads on the dynamics of elastic rods and elastic waves. This also includes important features such as parametric resonances and the dynamic loss of stability for the loads, whose magnitude is less than the classical Eulerian load. These important studies also extend to the non-linear regimes to analyse the growth of the post-critical deformations. 

Although most of micro-structured solids with defects are not periodic, they may possess structured interfaces or large subdomains, which include locally periodic patterns. Dynamic response of such elastic systems may be very unusual and sometimes counter-intuitive. The theory of Floquet-Bloch waves has been successfully used to explain the dynamic anisotropy, dispersion in multi-structured solids and localisation.       In particular, Slepyan \cite{Slepyan_2002} and Slepyan and Ryvkin \cite{Slepyan_Ryvkin_2010} have studied Floquet-Bloch waves in conjunction with the problems of dynamic fracture.
Structured interfaces, asymptotic analysis and the transmission problems for solids, containing structured interfaces, have been analysed by  Bigoni and Movchan \cite{Bigoni_2002}. For dynamic formulations in structured media where the long-wave approximations are not valid, the  high-frequency homogenisation is required, as explained in \cite{Craster_2014}, and this may also identify the anisotropy and localisation of waveforms in micro-structured solids. The so-called  band gap Green's functions, together with the dynamic defect modes, were  analysed in \cite{Movchan_Slepyan}.

In the previous papers by Piccolroaz and Movchan \cite{PM_2014} and Piccolroaz {\em et al.} \cite{PMC2017a}, Floquet-Bloch elastic waves were considered for elastic networks of elastic beams, with an additional rotational inertia. The authors also drew a connection between dynamic models of couple-stress elastic materials and structured Rayleigh beams. In particular, the Rayleigh beam model includes effects of the rotational inertia, that is excluded from the theory of the classical Euler-Bernoulli beams. It was also demonstrated that the rotational inertia is apparently significant for the high-frequency dynamic response, especially in problems involving the Dirac cone steering and analysis of the dispersion degeneracies.


\begin{figure}[!htb]
\centering
\includegraphics[width=120mm]{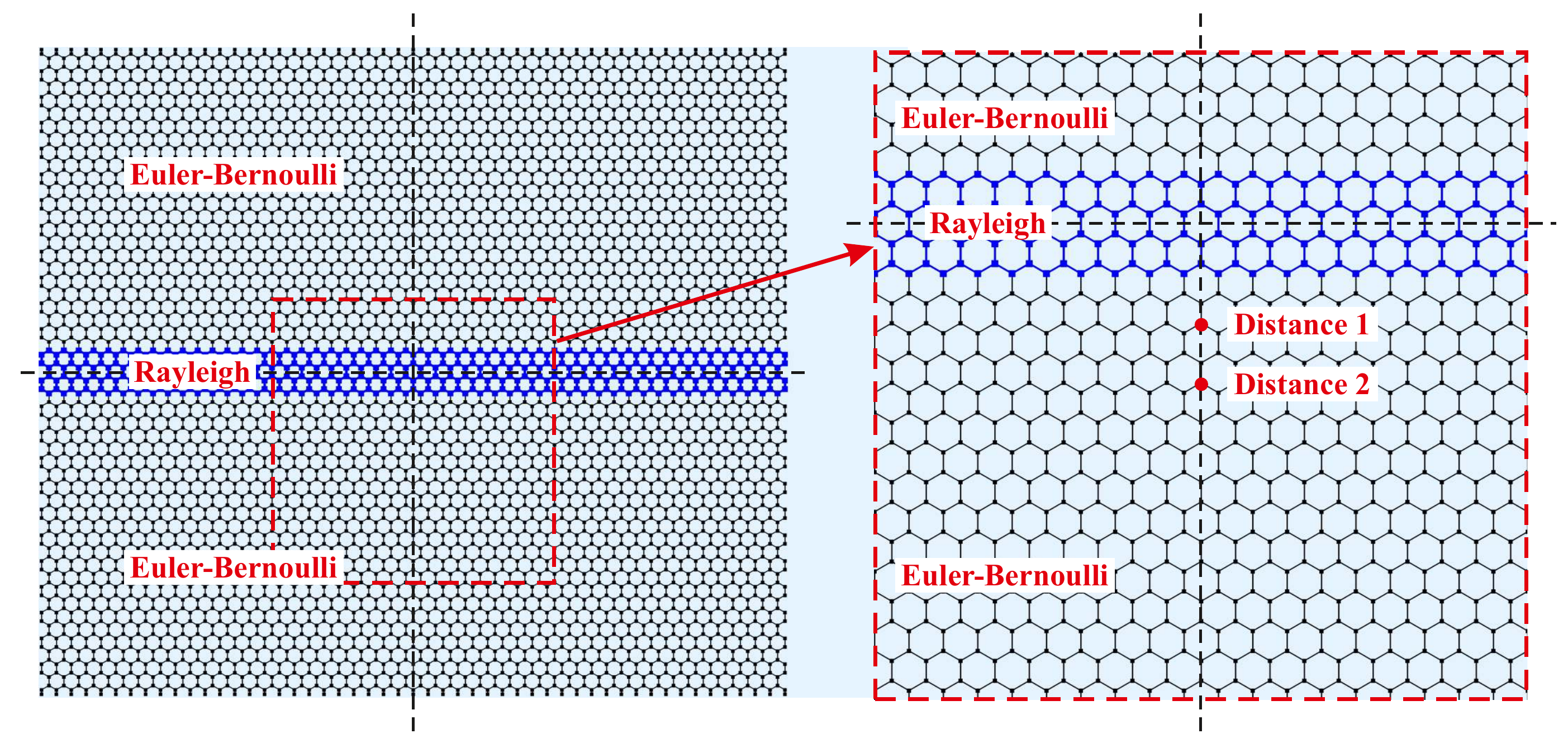}
\caption{\footnotesize A composite honeycomb lattice of flexural beams, which contains a structured interface consisting of the Rayleigh beams with an additional rotational inertia.}
\label{honeycomb_lattice_interface}
\end{figure}

Fig.~\ref{honeycomb_lattice_interface} shows a honeycomb flexural lattice containing a structured interface, where the classical Euler-Bernoulli beams are used to construct the ambient network, whereas the structured interface is built of the Rayleigh beams with an appropriately designed rotational inertia. 
When a point source is applied in the proximity of such a structured interface, it is demonstrated that the dynamic response of such a multi-scale solid may show localisation, negative refraction as well as dynamic anisotropy and interfacial edge waves.

The structure of the paper is as follows. The main definitions and the geometry of the Rayleigh beam network are given in Section 
\ref{sec02}. The elementary cell and the Bloch-Floquet junction conditions are discussed in Section \ref{BFjc}. 
Section \ref{sec03} is devoted to the dispersion of waves in the networks of the Euler-Bernoulli and of the Rayleigh beams. The dynamic anisotropy and the slowness contours corresponding to the dispersion diagrams are discussed in Section \ref{sec04}. The study of the forced vibrations and of the dynamic response of structured interfaces is presented in Section \ref{forced}. Finally, Section \ref{sec05} includes concluding remarks and discussion.

\section{The Rayleigh beam lattice.}
\label{sec02}

\begin{figure}[!htb]
\centering
\includegraphics[width=120mm]{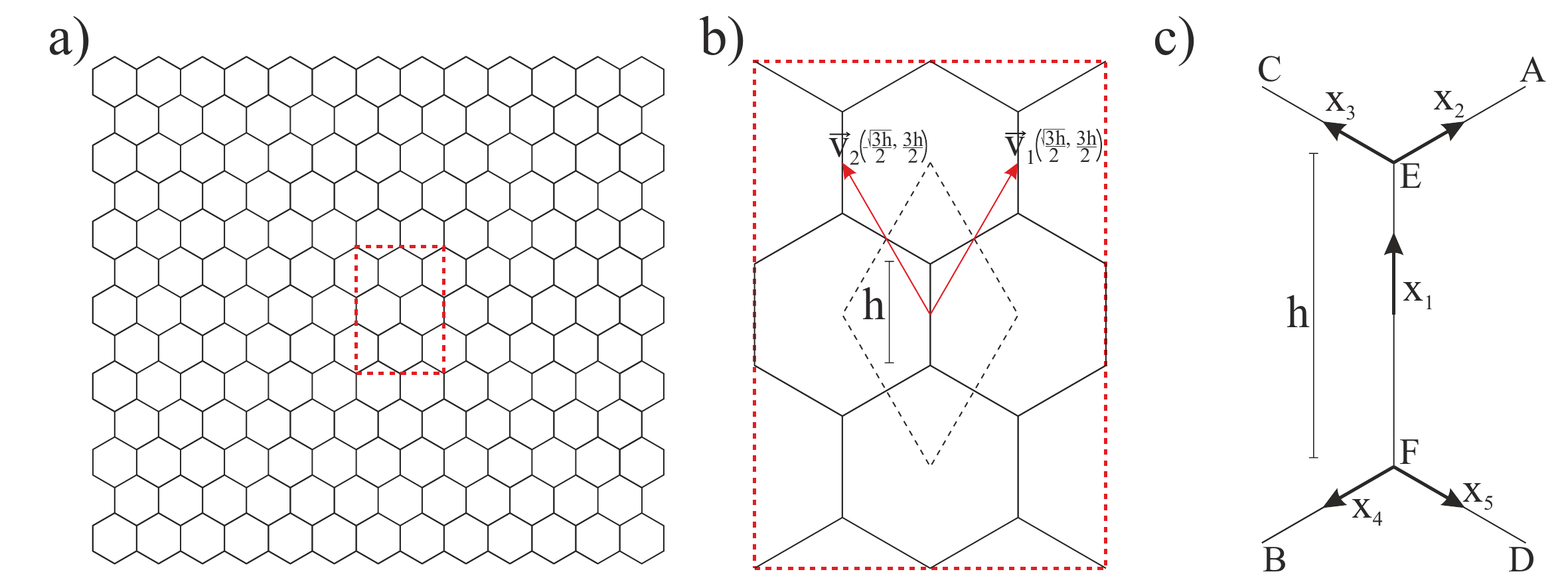}
\caption{\footnotesize  A periodic honeycomb lattice consisting of Rayleigh beams. The elementary cell, of a parallelogram shape is shown, together with the basis vectors of the lattice.}
\label{fig01}
\end{figure}

We begin by considering a doubly periodic honeycomb lattice consisting of the Rayleigh beams, which possess a rotational inertia. The geometry of the periodic structure is shown in Fig.~\ref{fig01}, which includes a periodic honeycomb lattice consisting of Rayleigh beams. The elementary cell, of a parallelogram shape is also shown, together with the basis vectors of the lattice, $\bv_1=(\sqrt{3}/2\ h, 3/2\ h)$ and $\bv_2=(-\sqrt{3}/2\ h, 3/2\ h)$ where $h$ is the length of the beams.

Within a one-dimensional ligament of the lattice, consider a time harmonic flexural wave, with the out-of-plane displacement 
$U(x, t) = u(x) \exp(i \omega t),$ with $x$ being a local spatial coordinate and $t$ denoting time. 

In the Rayleigh beam, the amplitude function $u(x)$ satisfies the governing equation
\begin{equation}
\label{eq:gov}
EI u''''(x) - (P - \rho I \omega^2) u'' + (\beta - \rho A \omega^2) u = 0
\end{equation}
Here the following notations are in use: $E$ is the Young modulus, $\beta$ is the stiffness of a Winkler type elastic foundation, $P$ the prestress, $\rho$ the mass density, $A$ the area of the cross-section, and $I$ the area moment of inertia of the cross-section.

Let $M$ and $V$ denote respectively the internal bending moment and the shear force, as follows: 
\begin{equation}
M(x) = -EI u''(x), \quad V(x) = -EI u'''(x) + (P - \rho I \omega^2) u'(x)
\end{equation}
If the Rayleigh beam is of infinite extent and prestress and elastic foundation are absent, then the solution  of (\ref{eq:gov}) can be written as
\begin{equation}
u(x) = \sum_{q=1}^{4} C_q e^{i \kappa_q x} \label{u_repr}
\end{equation}
where 
\begin{equation}
\kappa_{1,2,3,4} = \pm \frac{1}{r} \sqrt{\alpha\frac{R\omega^2}{2} \pm \sqrt{\alpha\frac{R^2\omega^4}{4} + R\omega^2}}
\label{kappa}
\end{equation}
The following notations are used here:
\begin{equation}
r = \sqrt{\frac{I}{A}}, \quad
R = \frac{\rho r^2}{E} = \frac{\rho I}{EA}
\end{equation}
Here $\alpha=1$ for the Rayleigh beams, and $\alpha=0$ for the Euler-Bernoulli beams. 

%
%
%
%
%

%
%
%


\section{The Floquet-Bloch and junction conditions within the elementary cell.}
\label{BFjc}

For the flexural lattice ligaments, shown in Fig.~\ref{fig01}, it is convenient to introduce local coordinates $x_j$ for each ligament and to refer to the junction points E and F, where three ligaments are connected.  

For the structure of five beams, shown in Fig.~\ref{fig01}(c), the local displacement amplitudes $u_q(x_q)$, $q= 1,\ldots,5,$  (there is no summation with respect to the repeated index $q$) are
\begin{equation}
u_q (x_q) = \sum_{p=1}^4 C_{pq} \exp(i \kappa_{pq} x_q), ~ q=1,\ldots, 5 \label{disp_nodal}
\end{equation}
and the quantities $\kappa_{pq}$ are defined by (\ref{kappa}) for each flexural ligament of the elementary cell. 
Hence, 
there are 20 constants, $C_{pq}$, $p=1,\cdots,4$, $q=1,\cdots,5$, which can be found by solving the system of equations derived from the 8 Floquet-Bloch conditions at the boundary of the unit cell (points A, B, C, D), and from the 12 junction conditions at the points E and F) .

In particular, for the points $A-B$ of the unit cell, the quasi-periodic conditions at $x_2 = h/2$ and $x_4 = h/2$ hold
\begin{equation}
u_2\left(\frac{h}{2}\right) = u_4\left(\frac{h}{2}\right) e^{ih(\sqrt{3}/2k_x+3/2k_y)} \label{eq6}
\end{equation}
\begin{equation}
u_2'\left(\frac{h}{2}\right) = -u_4'\left(\frac{h}{2}\right) e^{ih(\sqrt{3}/2k_x+3/2k_y)}
\end{equation}
\begin{equation}
- EI u_2''\left(\frac{h}{2}\right) = - EI u_4''\left(\frac{h}{2}\right) e^{ih(\sqrt{3}/2k_x+3/2k_y)}
\end{equation}
\begin{equation}
-EI u_2'''\left(\frac{h}{2}\right) - \rho I \omega^2 u_2'\left(\frac{h}{2}\right) =
- \left[ -EI u_4'''\left(\frac{h}{2}\right) - \rho I \omega^2 u_4'\left(\frac{h}{2}\right) \right] e^{ih(\sqrt{3}/2k_x+3/2k_y)}
\end{equation}
which prescribe the Floquet-Bloch shift across the unit cell for flexural displacement, rotation, internal moment and internal shear force. 

Analogous quasi-periodic boundary conditions apply to the points $C-D$ of the unit cell,
\begin{equation}
u_3\left(\frac{h}{2}\right) = u_5\left(\frac{h}{2}\right) e^{ih(-\sqrt{3}/2k_x+3/2k_y)}
\end{equation}
\begin{equation}
u_3'\left(\frac{h}{2}\right) = -u_5'\left(\frac{h}{2}\right) e^{ih(-\sqrt{3}/2k_x+3/2k_y)}
\end{equation}
\begin{equation}
- EI u_3''\left(\frac{h}{2}\right) = - EI u_5''\left(\frac{h}{2}\right) e^{ih(-\sqrt{3}/2k_x+3/2k_y)}
\end{equation}
\begin{equation}
-EI u_3'''\left(\frac{h}{2}\right) - \rho I \omega^2 u_3'\left(\frac{h}{2}\right) =
- \left[ -EI u_5'''\left(\frac{h}{2}\right)- \rho I \omega^2 u_5'\left(\frac{h}{2}\right) \right] e^{ih(-\sqrt{3}/2k_x+3/2k_y)}
\end{equation}
The junction conditions at the node $E$ are as follows.
Continuity of displacement implies
\begin{equation}
u_2(0) = u_1\left(\frac{h}{2}\right)
\end{equation}
\begin{equation}
u_3(0) = u_1\left(\frac{h}{2}\right)
\end{equation}
and the continuity of rotations yields
\begin{equation}
u_1'\left(\frac{h}{2}\right) = u_2'(0)+u_3'(0)
\end{equation}
equations of motion for the node $E$ become
\begin{multline}
\left[ -EI u_1'''\left(\frac{h}{2}\right)- \rho I \omega^2 u_1'\left(\frac{h}{2}\right) \right]=
\left[ -EI u_2'''(0) - \rho I \omega^2 u_2'(0) \right] +
\left[ -EI u_3'''(0) - \rho I \omega^2 u_3'(0) \right]
\end{multline}
\begin{equation}
EI u_1''\left(\frac{h}{2}\right)=\frac{EI}{2} u_2''(0)+\frac{EI}{2} u_3''(0)
\end{equation}
\begin{equation}
EI u_2''(0)=EI u_3''(0)
\end{equation}

\noindent
The junction conditions at the node $F$ include: continuity of displacement
\begin{equation}
u_4(0) = u_1\left(-\frac{h}{2}\right)
\end{equation}
\begin{equation}
u_5(0) = u_1\left(-\frac{h}{2}\right)
\end{equation}
continuity of rotations
\begin{equation}
u_1'\left(-\frac{h}{2}\right)+u_4'(0)+u_5'(0)=0
\end{equation}
and the equations of motion for the node $F$
\begin{multline}
\left[ -EI u_1'''\left(-\frac{h}{2}\right)- \rho I \omega^2 u_1'\left(-\frac{h}{2}\right) \right]=
\left[ -EI u_4'''(0) - \rho I \omega^2 u_4'(0) \right] +
\left[ -EI u_5'''(0) - \rho I \omega^2 u_5'(0) \right]
\end{multline}
\begin{equation}
EI u_1''\left(-\frac{h}{2}\right)=\frac{EI}{2} u_4''(0)+\frac{EI}{2} u_5''(0)
\end{equation}
\begin{equation}
EI u_4''(0)=EI u_5''(0) \label{eq25}
\end{equation}

\noindent
Equations (\ref{eq6})--(\ref{eq25}) comprise a homogeneous linear algebraic system for the 20 unknown constants, and the requirement of the vanishing of the determinant of the matrix of this algebraic system yields the dispersion equation for the Floquet-Bloch waves in the honeycomb periodic lattice of the Rayleigh beams. 

\section{Lower-dimensional model, dispersion properties.}
\label{sec03}

In the earlier paper \cite{PMC2017a}, which addressed the networks of flexural beams for square and rectangular lattices, it has been noted that the 
analysis of Floquet-Bloch waves contributes to the studies of the dynamic response of structured solids.
Here we analyse the dispersion properties of flexural waves in honeycomb systems with rotational inertia, and also make a comparison between periodic networks of different geometries.

\subsection{The algebraic system.}

Substitution of the representation 
\eq{disp_nodal} into the 20 relations 
(\ref{eq6})--(\ref{eq25}) 
leads to a system of linear algebraic equations with respect to the variables  $C_{pq}$. Introducing a vector $$\bC = \Big(C_{11}, C_{12}, C_{13}, C_{14}, \ldots, C_{51}, C_{52}, C_{53}, C_{54}\Big)$$ the matrix form of the equations is 
\beq
\bmA(\omega, \bk) \bC^T =0
\label{alg_syst}
\eeq
where $\bmA(\omega, \bk)$ is a $20 \times 20$ matrix  function, whose arguments are the radian frequency $\Go$ and the Bloch vector $\bk=(k_x,k_y)$. The dispersion equation has the form
\beq
\det \bmA(\omega, \bk) = 0
\label{disp_eq}
\eeq
The above equation defines implicitly the dispersion surfaces, and for any given $\bk$ from the Brillouin zone of the reciprocal lattice we identify a countable set of values of the spectral parameter $\Go$.
The dispersion equation has been solved and the results concerning the dispersion surfaces and, in particular, the Dirac cones and saddle points are discussed below.

\subsection{Dirac cones and standing waves.}

Honeycomb lattice of the Rayleigh beams has many attractive features, due to its high-order symmetry and possession of a dynamic response linked to dynamic anisotropy and to standing waves.

\begin{figure}[!htb]
\centering
\includegraphics[width=48mm]{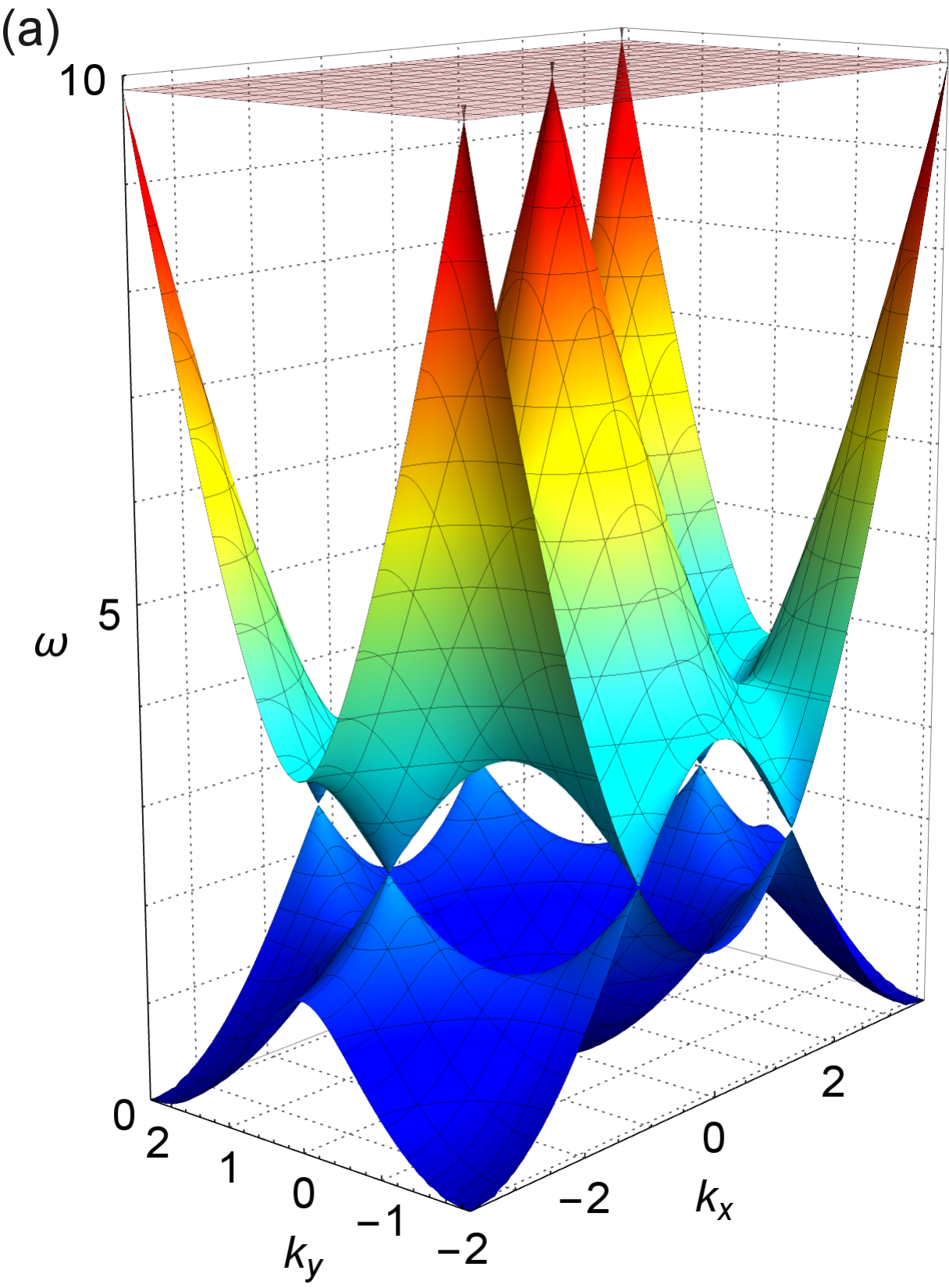}
\includegraphics[width=48mm]{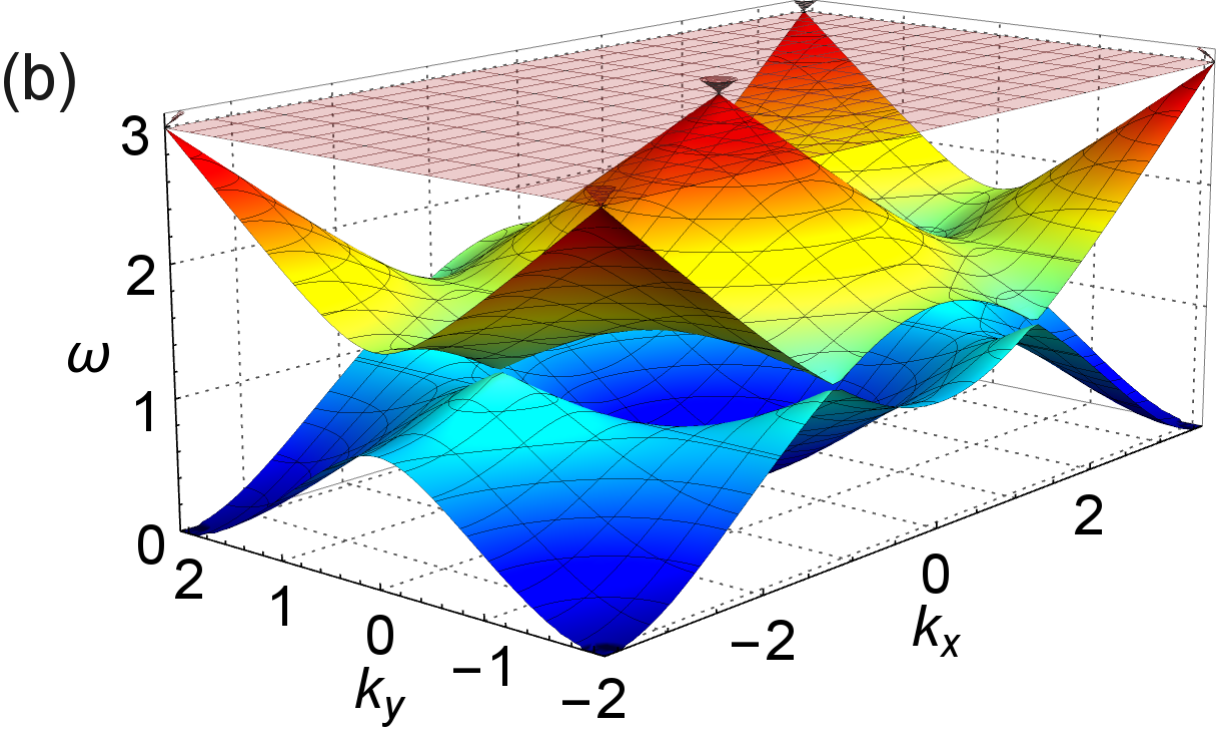}
\caption
{\footnotesize Dispersion surfaces for Floquet-Bloch waves in a periodic honeycomb lattice: (a)  the Euler-Bernoulli beam structure, (b) the Rayleigh beam structure. Neither prestress nor elastic foundations are present in this computation. The physical and geometrical parameters are chosen as follows: 
$E=1$, $\rho=1$, $A=1$, $I=1$, $h=1$.
The effect of the rotational inertia, inherent to the Rayleigh beams, is significant,   as the first several dispersion surfaces occur at much lower frequencies compared to the corresponding surfaces for the Euler-Bernoulli's beams as in part (a).
}
\label{fig3dh}
\end{figure}

\begin{figure}[!htb]
\centering
\includegraphics[width=48mm]{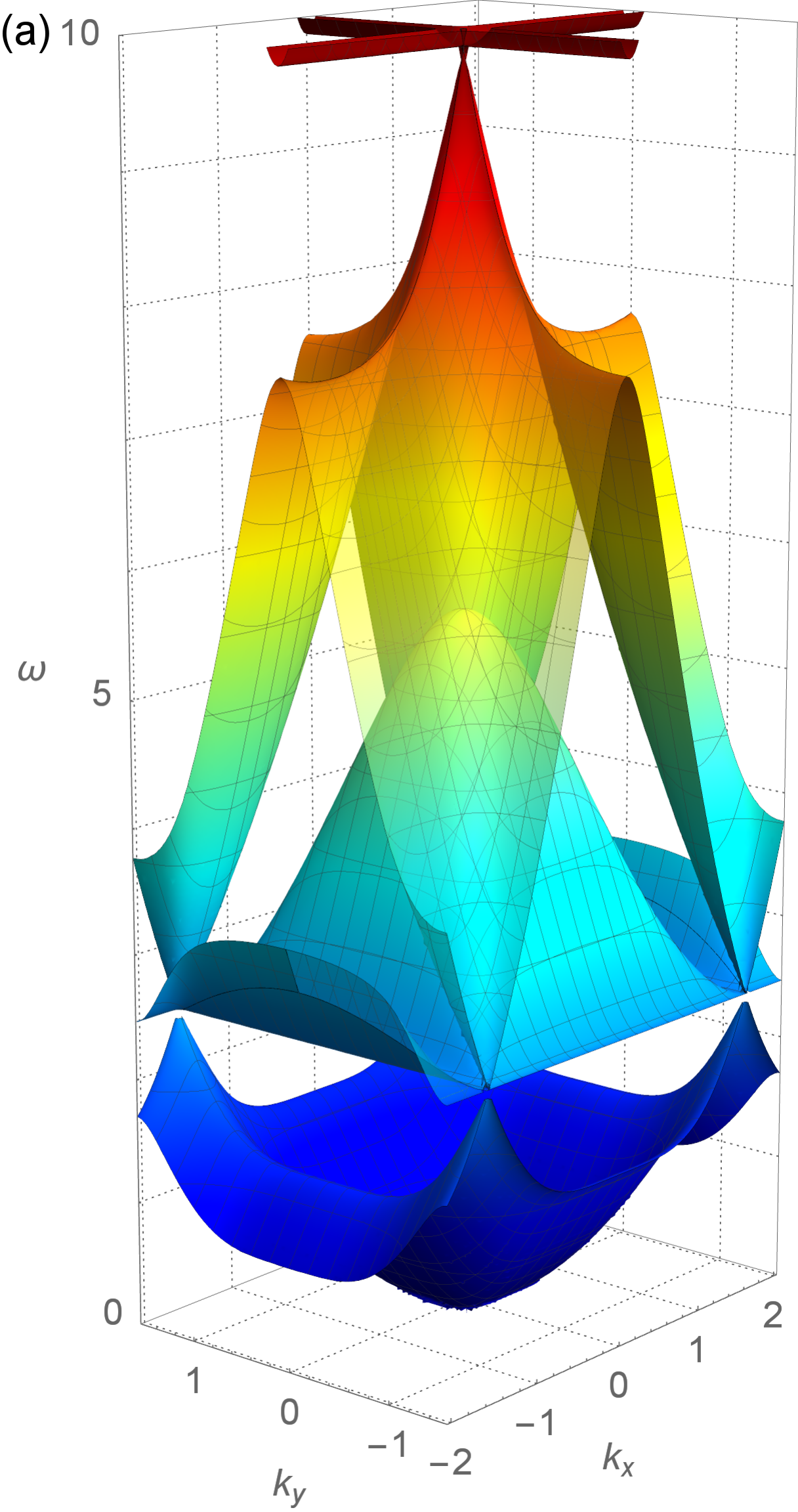}
\includegraphics[width=48mm]{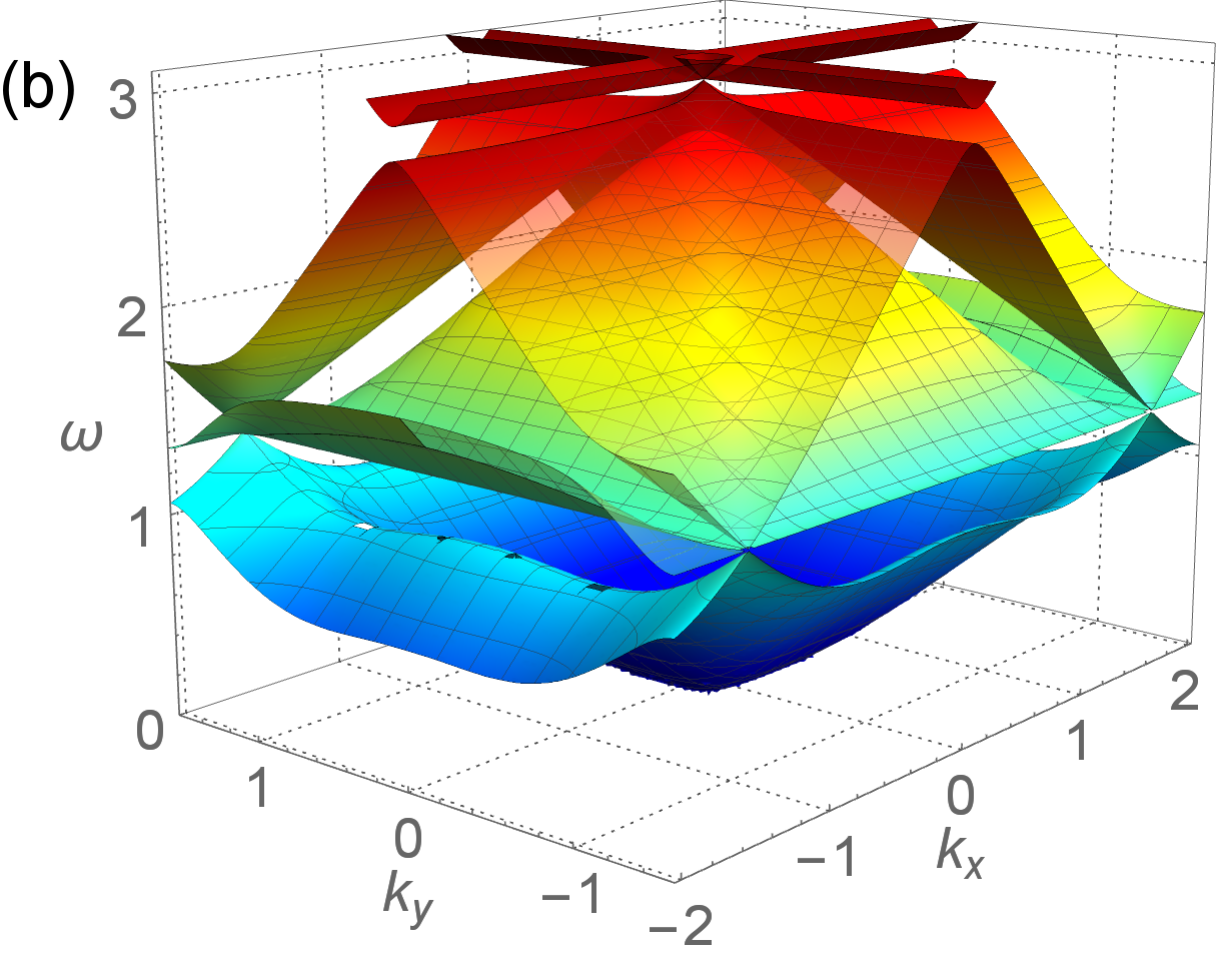}
\caption{\footnotesize The case of a square lattice. Dispersion surfaces are presented for 
the Euler-Bernoulli beam structure (a), and for the 
network of Rayleigh beams  (b). 
The parameter values, used in this combination, are $E=1$, $\rho=1$, $A=1$, $I=1$, ${h}=1$. 
} 
\label{fig3d}
\end{figure}

In particular, Figs.~\ref{fig3dh}, \ref{fig3d} shows the surfaces, which include a conical point, which corresponds to a finite non-zero frequency. The conical surfaces, adjacent to that point, are referred to as {\em Dirac cones}. It is also noted that a standing wave occurs, represented by a surface of zero slope traversing through the Dirac cone vertex.

Comparing the cases of the honeycomb networks of the Rayleigh beams, with an additional rotational inertia, and the classical case of the Euler-Bernoulli beams, we observe a clear distinction between the dispersion diagrams.
The first three dispersion surfaces for the Floquet-Bloch waves in the lattice, consisting of the Rayleigh beams, are placed at much lower frequencies compared to those for the Euler-Bernoulli beams,  as demonstrated in Fig.~\ref{fig3dh}. 

We also provide a comparison between the cases of different lattice geometries: the honeycomb lattice and the square lattice of flexural beams are discussed here. The results, concerning the dispersion of the Floquet-Bloch waves in the square flexural lattice were presented in the earlier work \cite{PMC2017a}. 
In particular, for the square lattice of elastic beams (both cases of Euler-Bernoulli and Rayleigh beams) the smoothness of dispersion surfaces and their degeneracies were investigated.
The effects of rotational inertia, attributed to the Rayleigh beams, deserve special attention, as illustrated in the text below.

For the honeycomb lattice, where the normalised physical and geometrical parameters are chosen as $E=1$, $\rho=1$, $A=1$, $I=1$, $h=1$, $P=0$, $\beta=0$, the dispersion equation, which relates the radian frequency $\omega$ and the wave vector $(k_x, k_y)$,  takes the form
\begin{multline}
\sin ({\Omega_1 h}) \sinh ({\Omega_2 h}) \left[4 \cos \left(\frac{\sqrt{3}}{2}{k_x h}-\frac{3}{2} {k_y h}\right) \right.
+4 \cos \left(\frac{\sqrt{3}}{2}{k_x h}+\frac{3}{2} {k_y h}\right) \\ \left.  +4 \cos \left(\sqrt{3} {k_x h}\right)-9 \cos (2 {\Omega_1 h})-3\right] 
\left[4 \cos \left(\frac{\sqrt{3}}{2}{k_x h}-\frac{3}{2} {k_y h}\right) \right. \\ \left. +4 \cos \left(\frac{\sqrt{3}}{2}{k_x h}+\frac{3}{2} {k_y h}\right)+4 \cos \left(\sqrt{3} {k_x h}\right)-9 \cos (2 {\Omega_2 h})-3\right]=0 \label{disp_honey}
\end{multline}
where
\begin{equation}
\Omega_1 = \frac{1}{r} \sqrt{\sqrt{\alpha\frac{R^2\omega^4}{4} + R\omega^2}+\alpha\frac{R\omega^2}{2}},
\quad
\Omega_2 = \frac{1}{r} \sqrt{\sqrt{\alpha\frac{R^2\omega^4}{4} + R\omega^2}-\alpha\frac{R\omega^2}{2}}
\end{equation}
in which $\alpha=1$ for Rayleigh beams and $\alpha=0$ for Euler-Bernoulli beams.

The high degree of symmetry of the honeycomb structure leads to a nicely factorised form of the dispersion equation, as above. In particular, two of the factors in the left-hand side are ${\bk}$-independent, and there exists a countable set of standing waves characterised by the flat dispersion surfaces at $\Omega_1(\omega) h = \pi k$ for any integer $k$.
One of such  flat surfaces is shown in Fig.~\ref{fig3dh}, together with the two dispersion surfaces that are not smooth and have conical points. This figure includes several parts: (a) the Euler-Bernoulli beam structure, (b) the Rayleigh beam structure. Neither prestress nor elastic foundations are present in this computation. The physical and geometrical parameters are chosen as follows: $E=1$, $\rho=1$, $A=1$, $I=1$, $h=1$. The effect of the rotational inertia, inherent to the Rayleigh beams, is significant,   as the first several dispersion surfaces occur at much lower frequencies compared to the corresponding surfaces for the Euler-Bernoulli's beams as in part (a).

Although a square lattice of flexural beams possesses interesting dispersion properties for the Floquet-Bloch waves, as shown in Fig.~\ref{fig3d}, there are significant differences compared to the case of the honeycomb flexural lattice. Dispersion surfaces are presented for the Euler-Bernoulli beam structure in part (a), and for the network of Rayleigh beams in part (b). The parameter values, used in this combination, are $E=1$, $\rho=1$, $A=1$, $I=1$, ${h}=1$.
In particular, the dispersion equation cannot be factorized to a similar form as in (\ref{disp_honey}), and the flat dispersion surfaces characterising standing waves are absent. Instead, the emphasis is made on the dynamic anisotropy, and special dispersion features of the Floquet-Bloch waves in the neighbourhoods of the vertices of the cones.  
 Following \cite{McPhedran_2015} we refer to these conical surfaces as ``Dirac cones'', and the radian frequency $\Go$ corresponding to the vertex of the cone is identified as a special resonant frequency, corresponding to a multiple root of the dispersion equation.

In particular, if the rotational inertia, the pre-stress and the stiffness of the elastic foundation are equal to zero, i.e.\ the Rayleigh beam becomes the classical Euler-Bernoulli beam, the dispersion equation \eq{disp_eq} for Floquet-Bloch waves in the square lattice takes the form
\begin{multline}
\sin \left(\Omega {h}\right)
\left[\cos (k_x {h}) + \cos (k_y {h}) - 2 \cos \left(\Omega {h}\right)\right]
\left[\cos (k_x {h}) - \cosh \left(\Omega {h}\right)\right]
\left[\cos (k_y {h}) - \cosh \left(\Omega {h}\right)\right] \\
-\sinh \left(\Omega {h}\right)
\left[\cos (k_x {h}) + \cos (k_y {h}) - 2 \cosh \left(\Omega {h}\right)\right]
\left[\cos (k_x {h}) - \cos \left(\Omega {h}\right)\right]
\left[\cos (k_y {h}) - \cos \left(\Omega {h}\right)\right] = 0
\end{multline}
where ${h}$ is the length of the ligaments in the square lattice, $\Omega = \sqrt{\omega} \sqrt[4]{\frac{\rho A}{EI}}$, and the corresponding dispersion diagram is shown in Fig.~\ref{fig3d}a.

\subsection{Resonant modes for elementary ligaments.}     
It turns out that the frequencies, corresponding to the Dirac cone vertices, and some of the corresponding vibration modes (i.e. standing waves) are identified as the natural frequencies and the eigenmodes of a simply supported beam, respectively. It also explains why when the size of beam ligaments and their physical properties are the same, the Dirac cone vertices for the honeycomb and for the square lattice occur at the same frequencies: 

\begin{equation}
\omega_n = \frac{n^2 \pi^2}{{h}^2} \sqrt{\frac{EI}{\rho A}} \qquad \text{for the Euler-Bernoulli beam} \label{swEB}
\end{equation}

\begin{equation}
\omega_n = \frac{n^2 \pi^2}{{h}} \sqrt{\frac{EI}{\rho (A{h}^2+n^2\pi^2I)}} \qquad \text{for the Rayleigh beam} \label{swR}
\end{equation}



The honeycomb lattice, even in the low-frequency regime, shows a very different dynamic response compared to the square lattice. It follows from the diagrams of Fig.~\ref{fig3dh} and Fig.~\ref{fig3d}, which present the dispersion surfaces for the Floquet-Bloch waves, that the honeycomb lattice is locally isotropic in the neighbourhood of the Dirac cone frequencies as well as at low frequencies. On the contrary, the orthotropy of the square lattice is apparent, and in particular, it leads to an unusual directional preference in the vicinity of the Dirac cones frequencies. 


Also the effect of the rotational inertia, which is present in the Rayleigh beam structure, becomes important, and consequently the dispersion surfaces representing structures consisting of the Rayleigh beams and structures consisting of the Euler-Bernoulli beams, become different, as seen in Figs.~\ref{fig3dh},~\ref{fig3d}.
Floquet-Bloch waves in a Rayleigh beam structure would show zero group velocity at much lower frequencies compared to the similar structure of the Euler-Bernoulli beams. At higher frequencies the Rayleigh beams show much richer behaviour, especially when it is concerned with the degeneracies and formation of the Dirac cones.

\begin{figure}[!htb]
\centering
\includegraphics[width=150mm]{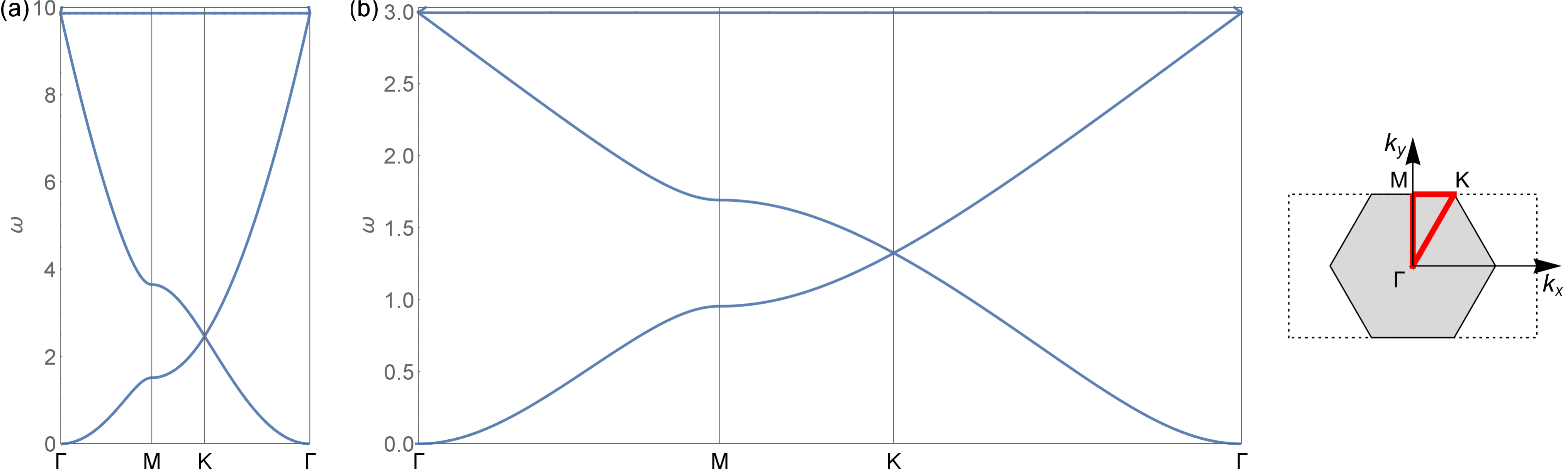}
\caption{
\footnotesize The cross-sectional dispersion diagrams along the boundary of the irreducible Brillouin zone, for Floquet-Bloch waves in the honeycomb lattice comprised of the Euler-Bernoulli beams (a) and the Rayleigh beams (b). 
{The inset on the right shows the contour $\GG M K \GG$ within the first Brillouin zone in the elementary cell of the reciprocal lattice; the dotted square corresponds to the computational window chosen to draw the dispersion surfaces in Fig.~\ref{fig3dh}.}
}
\label{figband2h}
\end{figure}

\begin{figure}[!htb]
\centering
\includegraphics[width=150mm]{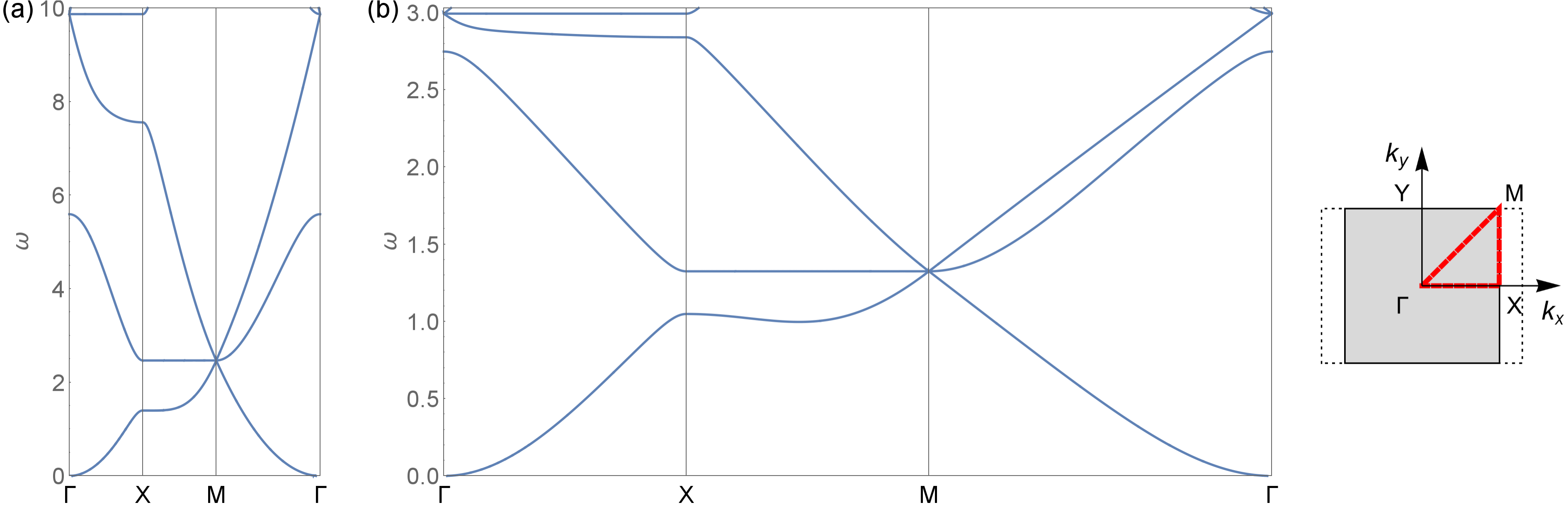}
\caption{
\footnotesize The cross-sectional dispersion diagrams along the boundary of the irreducible Brillouin zone, for Floquet-Bloch waves in the square networks of the Euler-Bernoulli beams (a) and the Rayleigh beams (b). 
{The inset on the right shows the contour $\GG X M \GG$ within the first Brillouin zone in the elementary cell of the reciprocal lattice; the dotted square corresponds to the computational window chosen to draw the dispersion surfaces in Fig.~\ref{fig3d}.}
}
\label{figband2}
\end{figure}

Figs.~\ref{figband2h}, \ref{figband2} complement the  dispersion surfaces diagrams by the cross-sectional plots along the boundaries of the irreducible  Brillouin zones in the reciprocal lattices constructed for the honeycomb and the square networks of flexural beams.
In particular, Fig.~\ref{figband2h} includes the cross-sectional dispersion diagrams along the boundary of the irreducible Brillouin zone, for Floquet-Bloch waves in the honeycomb lattice comprised of the Euler-Bernoulli beams in part (a) and the Rayleigh beams in part (b). 
{The inset on the right shows the contour $\GG M K \GG$ within the first Brillouin zone in the elementary cell of the reciprocal lattice; the dotted square corresponds to the computational window chosen to draw the dispersion surfaces in Fig.~\ref{fig3dh}.}

Similar diagrams for square networks are shown in Fig.~\ref{figband2} where the case of the Euler-Bernoulli beams is shown in part (a) and the Rayleigh beams case is displayed in part (b). 
{The inset on the right shows the contour $\GG X M \GG$ within the first Brillouin zone in the elementary cell of the reciprocal lattice; the dotted square corresponds to the computational window chosen to draw the dispersion surfaces in Fig.~\ref{fig3d}.

Special attention is given to the Dirac cones, represented by the intersecting dispersion curves. In particular, for the first intersection the case of the square lattice shows a triple root of the dispersion equation, with the standing wave being represented by the flat band crossing through the Dirac cone vertex. On the contrary, the Floquet-Bloch waves in the honeycomb network flexural waves also posses the Dirac cone mode, but for the first intersection depicted in Fig.~\ref{figband2h} the standing wave is absent. We note that the standing wave for honeycomb network of flexural beams occurs at the second intersection, and this mode is represented by the flat band at the frequency defined by formulae \eq{swEB},  \eq{swR}.
Also, we remark that the additional rotational inertia attributed to the Rayleigh beams leads to the dispersion curves being compressed towards the horizontal axis compared to those on the diagrams presented for the Euler-Bernoulli beams. This feature is clearly visible from the comparison of the parts (a) and (b) on the dispersion diagrams shown in Figs.~\ref{figband2h} and \ref{figband2}.



\section{Floquet-Bloch waves in a honeycomb network. Saddle points and slowness contours.}
\label{sec04}


The slowness contours (often referred to as isofrequency contours) are useful to identify stationary points on the dispersion surface as well as the dynamic anisotropy of the structured medium. %
A comprehensive  exposition of such an approach for flexural waves in periodic square lattices is discussed in \cite{PMC2017a}.

\begin{figure}[!htb]
\centering
\includegraphics[width=120mm]{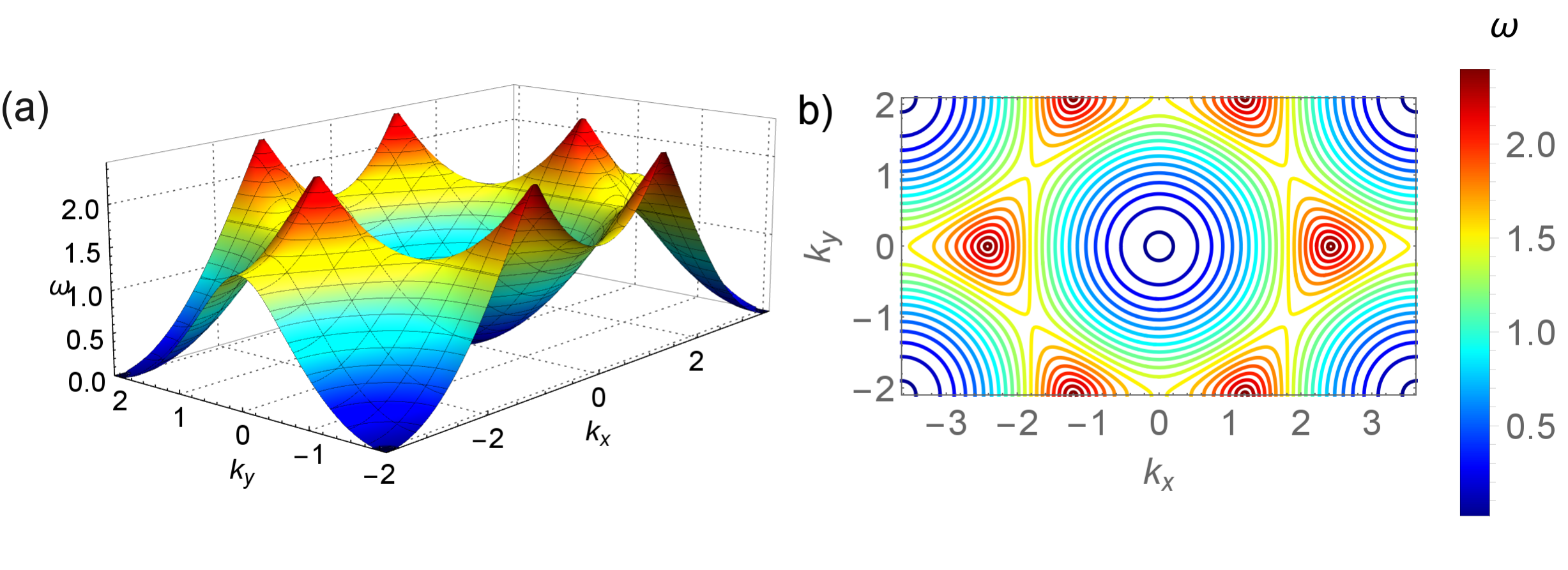} \\[6mm]
\includegraphics[width=120mm]{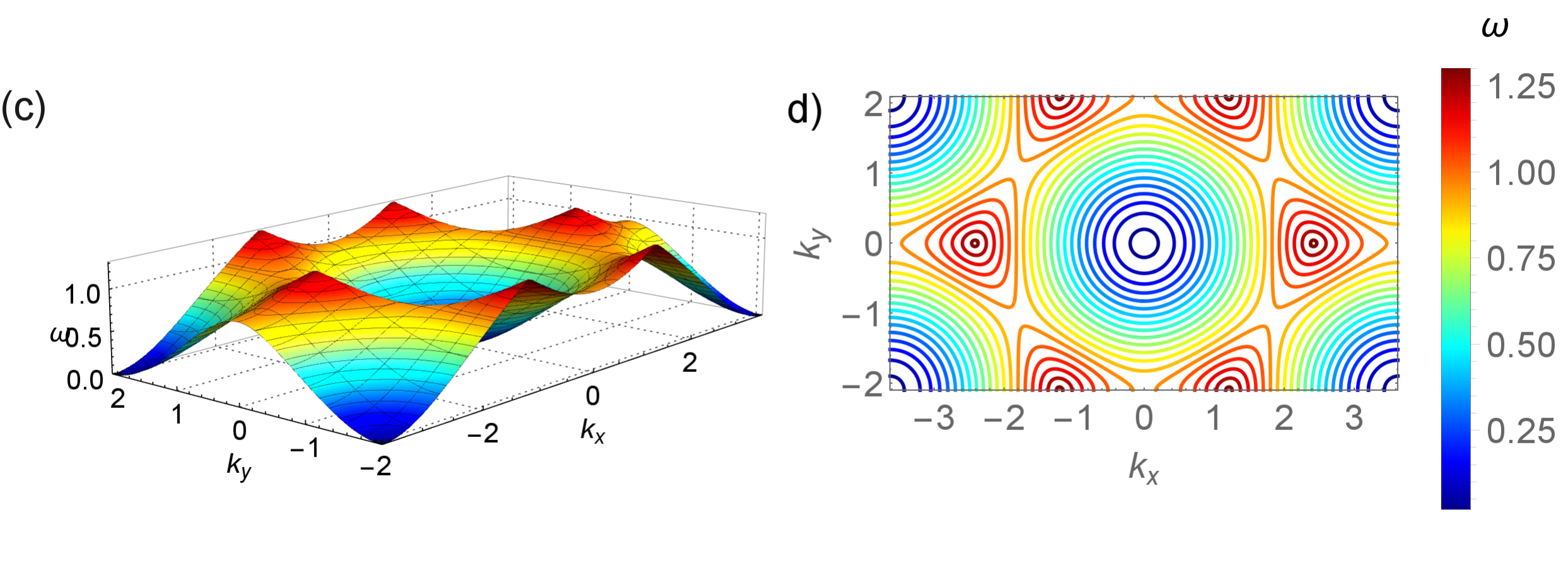}
\caption{
\footnotesize
First dispersion surface and the corresponding isofrequencies contours for the Euler-Bernoulli beam honeycomb lattice (parts (a) and (b)) and for the honeycomb lattice of the Rayleigh beams (parts (c) and (d)). The Dirac cone is shown in both configurations. The slowness contours around the origin bound non-convex domains in the diagram (d), for the Rayleigh beams, in contrast with the diagram (b), corresponding to the Euler-Bernoulli beams. 
}
\label{slowness_contours}
\end{figure}

\begin{figure}[!htb]
\centering
\includegraphics[width=120mm]{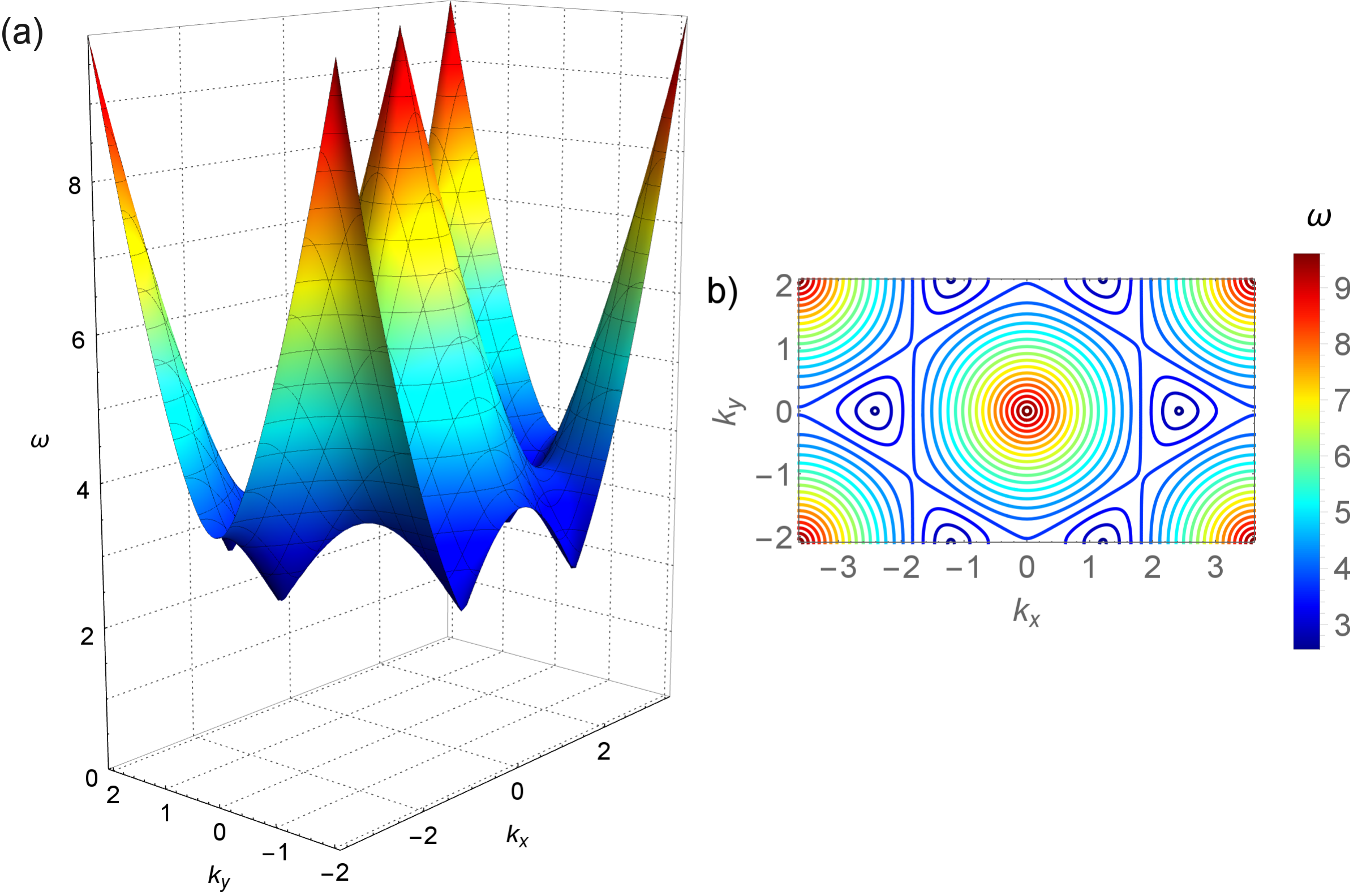} \\[6mm]
\includegraphics[width=120mm]{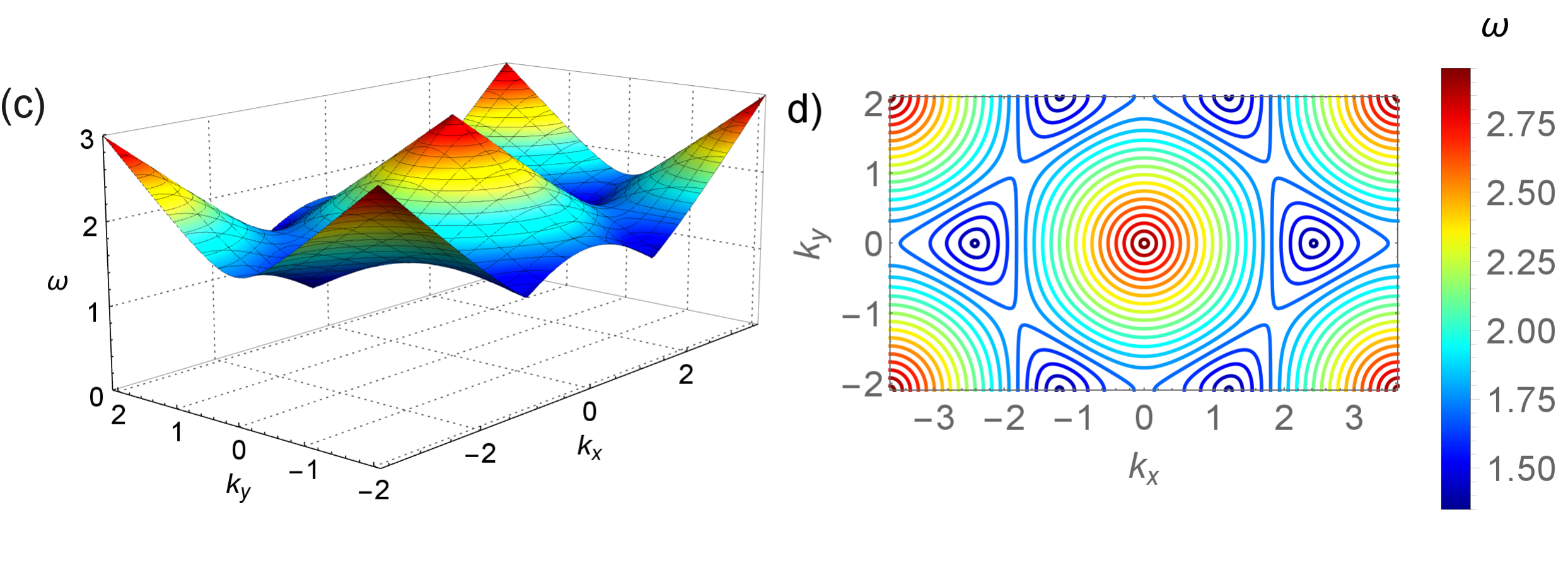}
\caption{\footnotesize
The second dispersion surface, including the Dirac cone, presented for the Floquet-Bloch waves in the case of the Euler-Bernoulli beam square lattice (parts (a) and (b)) and of the Rayleigh beam square lattice (parts (c) and (d)). Preferential directions along the coordinate axes are clearly identified.
}
\label{slowness_contours2}
\end{figure}

For Floquet-Bloch flexural waves in the honeycomb lattice, the dispersion surfaces and the corresponding slowness contours, are presented in Figs.~\ref{slowness_contours}, \ref{slowness_contours2} for the cases of the Rayleigh beams and the Euler-Bernoulli beams. 
In particular, the diagrams in Fig.~\ref{slowness_contours} represent the first dispersion surface (so-called ``acoustic band'') for the networks of Euler-Bernoulli beams and of the Rayleigh beams, whereas the diagrams in  Fig.~\ref{slowness_contours2} correspond to the second dispersion surface (the ``optical band'') for the networks of Euler-Bernoulli beams and of the Rayleigh beams. The cases of the Euler-Bernoulli beam honeycomb lattice correspond to parts (a) and (b) and the computations for the honeycomb lattice of the Rayleigh beams are presented in parts (c) and (d). 

These dispersion surfaces show the presence of the conical points (vertices of the Dirac cones) as well as the saddle points. Locally isotropic pattern is observed in the neighbourhood of the Dirac cones vertices, whereas a strong dynamic anisotropy is featured at frequencies corresponding to the saddle points.

The additional rotational inertia, attributed to the case of the Rayleigh beams, leads to lower frequencies of the Dirac cone vertices and the saddle points compared to similar networks of the Euler-Bernoulli beams.

In the following section, we give an illustration of the dynamic response of the honeycomb flexural lattice. In particular, we consider the case of a structured interface built of the Rayleigh beams embedded into the ambient geometrically identical lattice comprised of the Euler-Bernoulli beams, as illustrated in Fig.~\ref{honeycomb_lattice_interface}.


\section{Forced vibrations of a honeycomb flexural network.}
\label{forced}


%


Here we present the results of the finite element  simulations for the Euler-Bernoulli and for the Rayleigh beams programmed in COMSOL Multiphysics for a honeycomb network subjected to a time-harmonic transverse point force or several point forces applied at the lattice junctions.
The forces are applied in the direction perpendicular to the $(x, y)-$plane.

To simulate a dynamic response of an infinite lattice with a finite-size computational window and to avoid wave reflection at the boundaries, Rayleigh and Euler-Bernoulli beams with damping were also programmed and introduced at the boundary of the computational domain. This was achieved by replacing the Young modulus $E$ by a complex value, $E(1+i\eta)$. These beams were used to build a damping layer around the perimeter of the finite-size lattice, and the viscous parameter $\eta$ was chosen so that to suppress the wave reflection.

\subsection{A homogeneous flexural network of Rayleigh beams.}

In Fig.~\ref{fig_homo_RA} we consider a uniform network where six identical time-harmonic point forces are applied at the junctions of the hexagonal cell.

\begin{figure}[!htb]
\centering
\includegraphics[width=160mm]{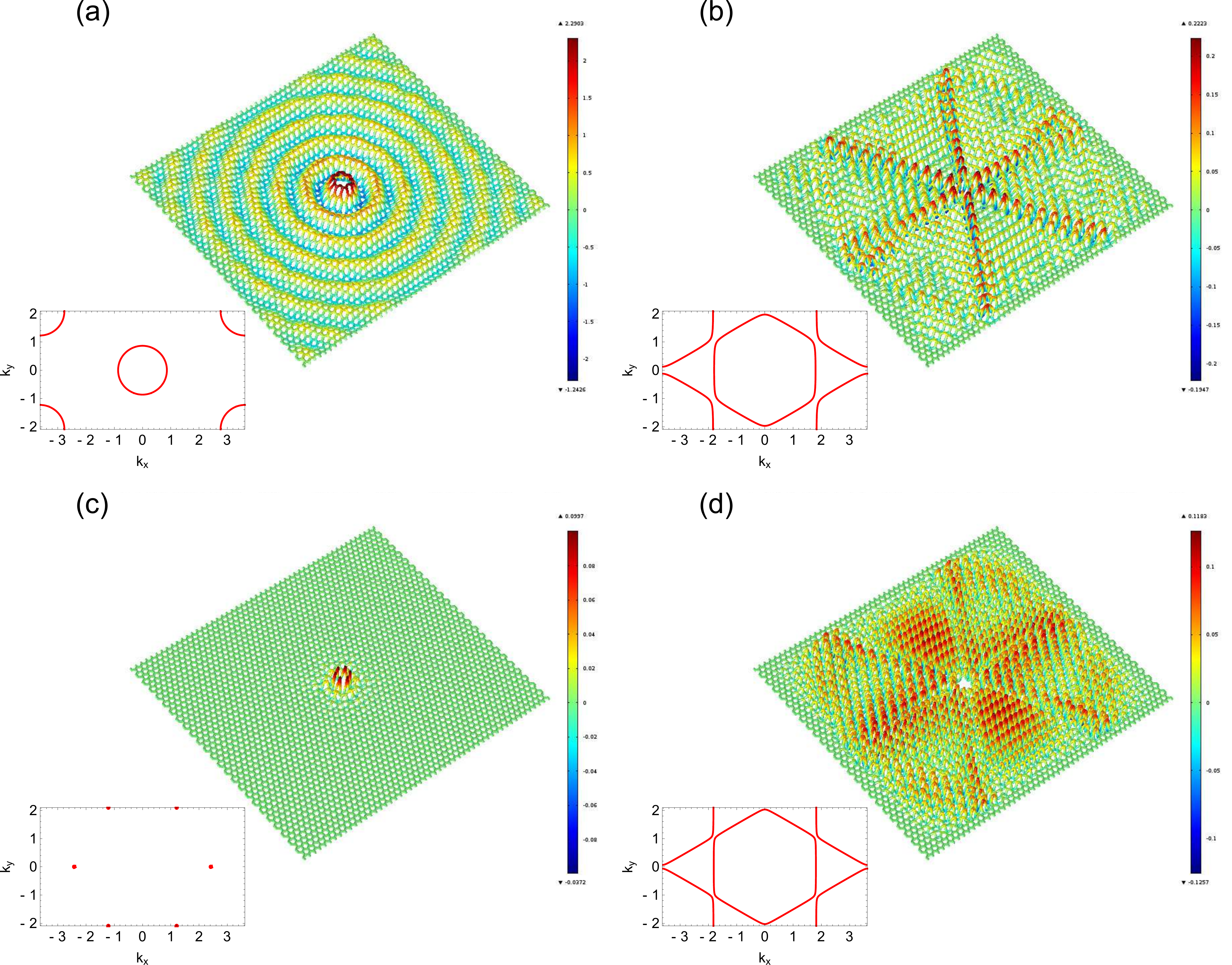}
\caption{
\footnotesize
The field patterns in the network of the Rayleigh beams, where  forced vibrations are generated by six time-harmonic identical forces applied at the junction points of the hexagonal cell.
(a) The radian frequency $\Go = 2 \pi f = 0.314$ is sufficiently low, and the dynamic response appears to be isotropic. (b) The radian frequency $\Go = 0.942$ is in the neighbourhood of the first saddle point frequency, where strong dynamic anisotropy is observed. (c) The radian frequency $\Go = 1.319$ is in the neighbourhood of the Dirac cone vertex, and the waveform is localised. (d) Another group of saddle points occurs at the radian frequency of  $\Go = 1.696$, where a strong dynamic anisotropy is observed. Here and in the following figures, inserts represent slowness contour diagrams for Floquet-Bloch waves at the given frequencies.
}
\label{fig_homo_RA}
\end{figure}

Several field patterns are observed for different values of the input frequency.
Namely, when the frequency is sufficiently low, as in part (a) of Fig.~\ref{fig_homo_RA}, the radial wave pattern appears to be isotropic, as expected in the low-frequency regime for the honeycomb lattice.
However, the increase of the input frequency leads to radical, but predictable  changes in the frequency response of the network of the Rayleigh beams.  
In the cases (b) and (d) of Fig.~\ref{fig_homo_RA}, the frequency values are close to those of the saddle points on the dispersion surfaces, and hence strong dynamic anisotropy is observed. On the contrary, strong localisation shown in the case (c) corresponds to the vertex of the Dirac cone. 
The frequency values are chosen as follows. (a) The radian frequency $\Go = 2 \pi f = 0.314$ is sufficiently low, and the dynamic response appears to be isotropic. (b) The radian frequency $\Go = 0.942$ is in the neighbourhood of the first saddle point frequency, where strong dynamic anisotropy is observed. (c) The radian frequency $\Go = 1.319$ is in the neighbourhood of the Dirac cone vertex, and the waveform is localised. (d) Another group of saddle points occurs at the radian frequency of  $\Go = 1.696$, where a strong dynamic anisotropy is observed. Here and in the following figures, inserts represent slowness contour diagrams for Floquet-Bloch waves at the given frequencies.

\subsection{Structured interface possessing rotational inertia.}

Here we consider a geometrically homogeneous lattice of flexural beams, but we assume that within a layer of finite thickness the classical Euler-Bernoulli beams are replaced by the Rayleigh beams, which possess a rotational inertia, as described by the governing equation  \eq{eq:gov}.
A point force excitation is applied at a junction, outside the structured layer, as depicted in Fig.~\ref{honeycomb_lattice_interface}. 
In different frequency regimes, the dynamic response of the structured layer is investigated here.

First, in Fig.~\ref{fig_inter1} we present two cases, which include the low frequency excitation (part (a)) and an excitation at a higher frequency (part (b)), where a negative refraction is observed.

\begin{figure}[!htb]
\centering
\includegraphics[width=150mm]{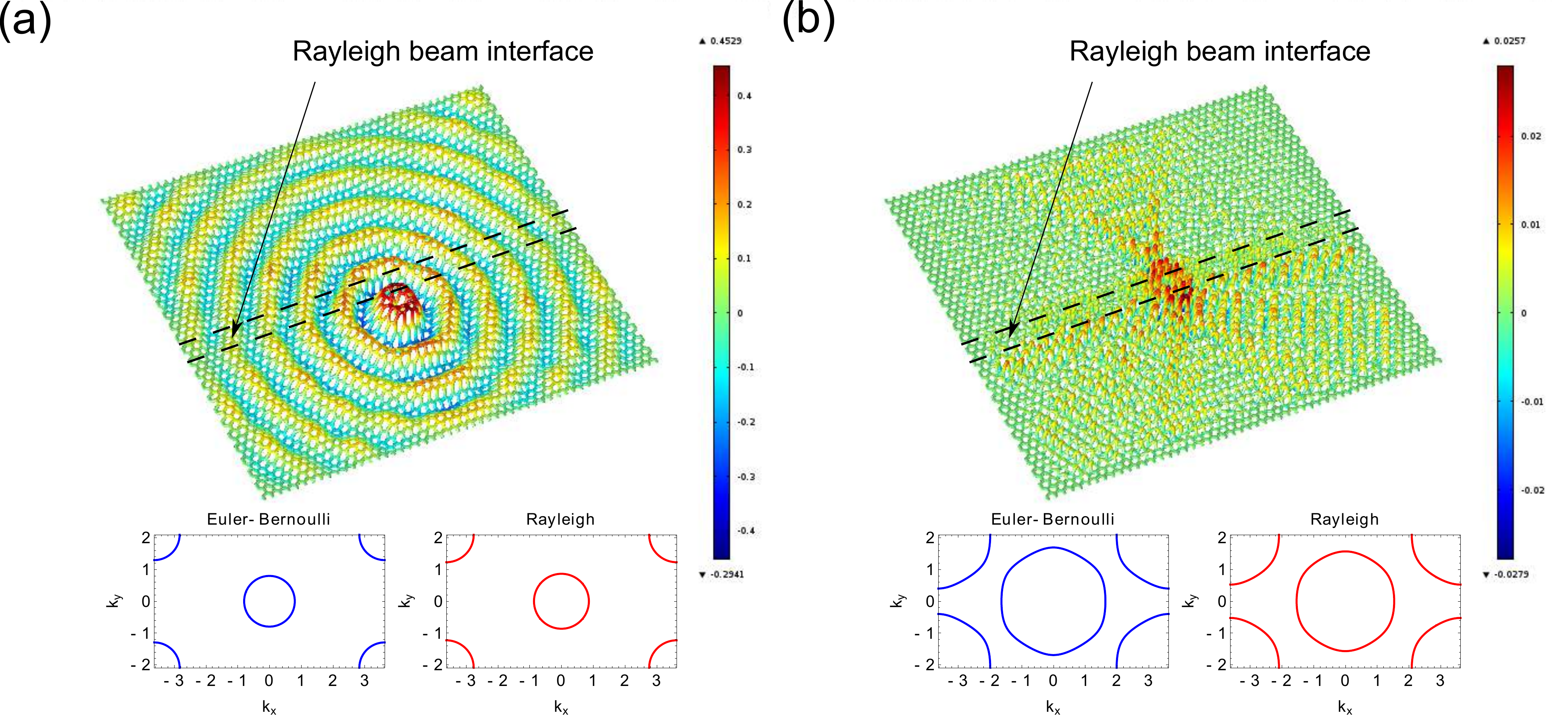}
\caption{
\footnotesize
A dynamic response of the structured layer, built of the Rayleigh beams. The ambient lattice, surrounding the structured interface, consists of the classical Euler-Bernoulli beams.      The field patterns are presented for the two cases, where a point force is applied at the junction separated from the structured interface by the distance equal to the double diameter of the elementary cell of the honeycomb lattice (``distance 2''), as shown in Fig.~\ref{honeycomb_lattice_interface}.
(a) A sufficiently low radian frequency $\Go = 2 \pi f = 0.314$ corresponds to the isotropic response, where the action of the interface layer is negligibly small. (b) The normalised radian frequency $\Go = 4.082$ corresponds to the case of Floquet-Bloch waves, which occur near saddle points on the dispersion diagrams for both the Euler-Bernoulli and the Rayleigh beams.  In this case, the rotational inertia of the interface layer becomes important, as it leads to the negative refraction and consequently focussing of the elastic flexural wave across the interface.
}
\label{fig_inter1}
\end{figure}

This is understandable, as in the low frequency regime the Euler-Bernoulli and the Rayleigh beams appear to be very similar in terms of their dynamic response to an external load, and hence the structured interface is ``invisible''.
On the other hand, as the frequency increases, the corresponding dispersion diagrams in Fig.~\ref{fig3dh} for the Floquet-Bloch waves  show the presence of the saddle points as well the Dirac cones. Consequently, this leads to an interesting dynamic response of the structured interface:
at the normalised radian frequency of $\Go = 4.082$, we observe saddle points on the dispersion diagrams for both the Euler-Bernoulli and for the Rayleigh beam networks. Such a regime implies a strong dynamic anisotropy, and consequently a negative refraction and focussing by the structured interface are observed in this simulation.
The field patterns are presented for the two cases, where a point force is applied at the junction separated from the structured interface by the distance equal to the double diameter of the elementary cell of the honeycomb lattice (``distance 2''), as shown in Fig.~\ref{honeycomb_lattice_interface}.
(a) A sufficiently low radian frequency $\Go = 2 \pi f = 0.314$ corresponds to the isotropic response, where the action of the interface layer is negligibly small. (b) The normalised radian frequency $\Go = 4.082$ corresponds to the case of Floquet-Bloch waves, which occur near saddle points on the dispersion diagrams for both the Euler-Bernoulli and the Rayleigh beams.  In this case, the rotational inertia of the interface layer becomes important, as it leads to the negative refraction and consequently focussing of the elastic flexural wave across the interface.

Furthermore, another Fig.~\ref{fig_inter2} displays the field plots and shows examples of strong dynamic anisotropy as well localisation within the structured interface consisting of the Rayleigh beams. The time-harmonic point force is placed at the distance 1 from the interface, as illustrated in Fig.~\ref{honeycomb_lattice_interface}. 
In particular, the part (a) of  Fig.~\ref{fig_inter2} shows the localisation within the structured interface, parts (c) and (d) show the edge-wave modes, and the field pattern of the part (b) shows the wave blockage by the structured interface and a strong dynamic anisotropy exhibited by the ambient lattice of the Euler-Bernoulli beams.      
The field patterns are presented for four cases, where a point force is applied at the junction separated from the structured interface by the distance equal to a diameter of the elementary cell of the honeycomb lattice (``distance 1''), as shown in Fig.~\ref{honeycomb_lattice_interface}.
(a) The normalised radian frequency $\Go = 2 \pi f = 1.005$ corresponds to a frequency below the first saddle point for the ambient lattice, where the dynamic response is almost isotropic, but the same frequency corresponds to the saddle point of the Rayleigh beam network used in the structured interface; hence the strong anisotropy is shown within the structured interface. (b) The normalised radian frequency $\Go = 1.508$  corresponds to the saddle point on the dispersion diagram for the Floquet-Bloch waves in the ambient lattice, and the same frequency corresponds to the neighbourhood of the Dirac cone vertex for the network of the Rayleigh beams. (c) The normalised radian frequency $\Go = 2.513$ corresponds to the neighbourhoods of the Dirac cone vertices for both Euler-Bernoulli and the Rayleigh beams on the first dispersion surface and the second dispersion surface, respectively  (d)  The forced excitation at the normalised radian frequency $\Go = 9.111$ gives a waveguide vibration mode, and it is shown that this regime is close to the Dirac cones of different orders for the Euler-Bernoulli and the Rayleigh beam networks, respectively.

\begin{figure}[!htb]
\centering
\includegraphics[width=150mm]{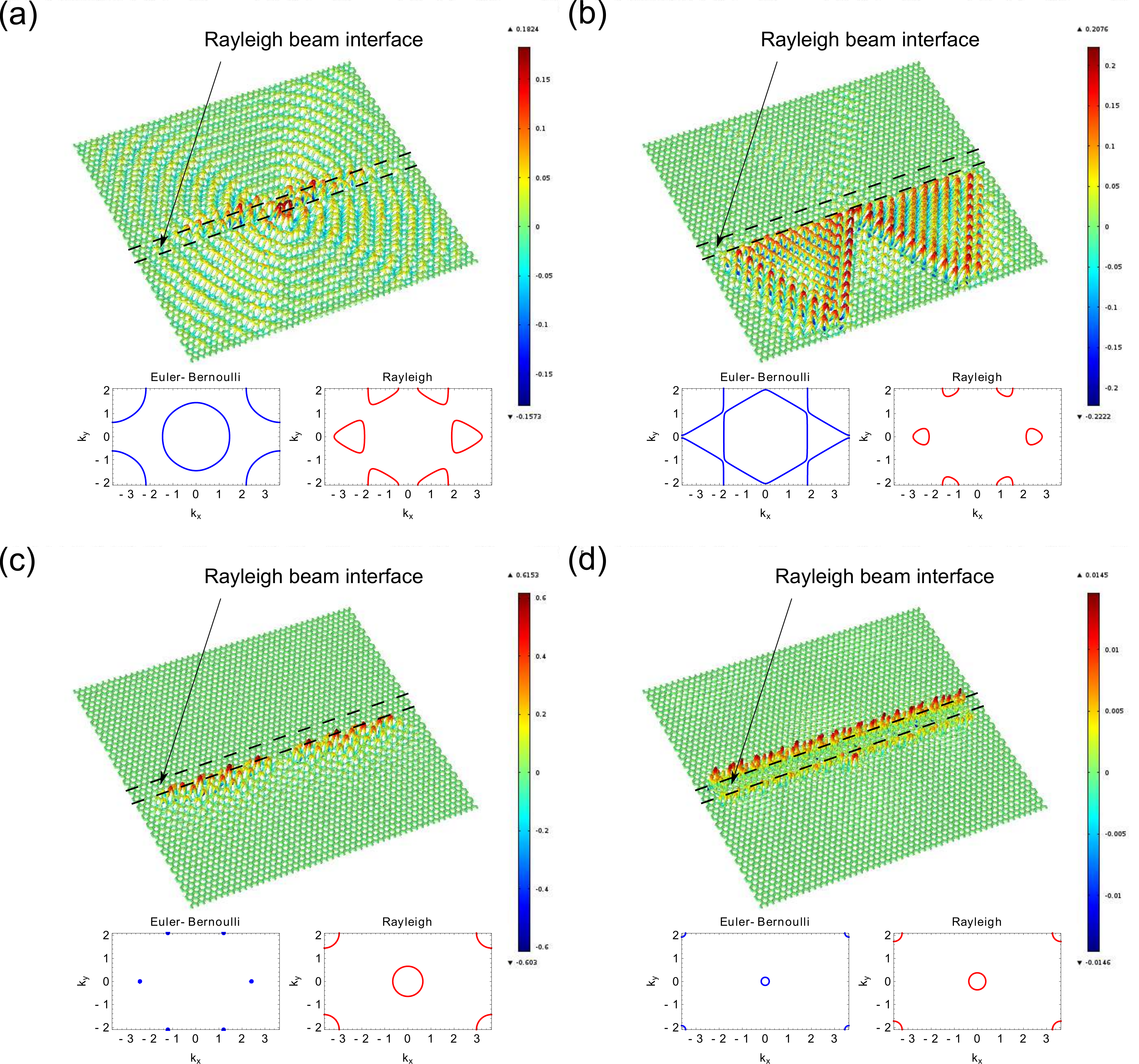}
\caption{
\footnotesize
A dynamic response of the structured layer, built of the Rayleigh beams. The ambient lattice, surrounding the structured interface, consists of the classical Euler-Bernoulli beams.      The field patterns are presented for four cases, where a point force is applied at the junction separated from the structured interface by the distance equal to a diameter of the elementary cell of the honeycomb lattice (``distance 1''), as shown in Fig.~\ref{honeycomb_lattice_interface}.
(a) The normalised radian frequency $\Go = 2 \pi f = 1.005$ corresponds to a frequency below the first saddle point for the ambient lattice, where the dynamic response is almost isotropic, but the same frequency corresponds to the saddle point of the Rayleigh beam network used in the structured interface; hence the strong anisotropy is shown within the structured interface. (b) The normalised radian frequency $\Go = 1.508$  corresponds to the saddle point on the dispersion diagram for the Floquet-Bloch waves in the ambient lattice, and the same frequency corresponds to the neighbourhood of the Dirac cone vertex for the network of the Rayleigh beams. (c) The normalised radian frequency $\Go = 2.513$ corresponds to the neighbourhoods of the Dirac cone vertices for both Euler-Bernoulli and the Rayleigh beams on the first dispersion surface and the second dispersion surface, respectively  (d)  The forced excitation at the normalised radian frequency $\Go = 9.111$ gives a waveguide vibration mode, and it is shown that this regime is close to the Dirac cones of different orders for the Euler-Bernoulli and the Rayleigh beam networks, respectively.
}
\label{fig_inter2}
\end{figure}






\section{Conclusion.}
\label{sec05}

The modelling of the beam networks which possess rotational inertia has revealed novel and counter-intuitive properties related to the dynamic response of the multi-scale solids.
Specifically, we have focused our attention on the regimes, which correspond to the saddle points or neighbourhoods of the Dirac cones on the dispersion diagrams, constructed for the Floquet-Bloch waves in the periodic networks of the Rayleigh or the Euler-Bernoulli beams.
The waveforms, and in particular standing waves also depend on the geometry of the periodic network, and we have given the comparative outline for the cases of square and honeycomb lattices, with the emphasis on the dynamic anisotropy and localisation.
Closed form asymptotic estimates for frequencies corresponding to the vertices of the Dirac cones and of the standing waves provide an additional useful tool, which is essential in problems of optimal design of phononic filters and polarisers of elastic waves.
Finally, the dynamic response  of the structured interfaces with an additional rotational inertia has been investigated, with the emphasis on edge waves, negative refraction and wave trapping.
This work naturally extends to the area of metamaterials design, and to control of flexural waves by multi-scale elastic networks.

\vspace{6mm}
{\bf Acknowledgements}. AP would like to acknowledge financial support from the
European Union's Seventh Framework Programme FP7/2007-2013/ under REA grant
agreement number PCIG13-GA-2013-618375-MeMic. 
Major part of the work was carried out while AM was visiting the University of Trento in 2016 with the support from the
European Union's Grant ERC-2013-ADG-340561-INSTABILITIES, which is gratefully acknowledged.
AM also acknowledges the support from the UK EPSRC Program Grant  EP/L024926/1.
LC acknowledges financial support from the University of Trento, within the research project 2014 entitled ``3D printed metallic foams for biomedical applications: understanding and improving their mechanical behavior''.

\bibliographystyle{jabbrv_unsrt}
\bibliography
{%
roaz1}

%
%

\end{document}

%% file: definitions_ijss.tex
\newcommand{\beq}{\begin{equation}}
\newcommand{\eeq}{\end{equation}}

\newcommand{\beqar}{\begin{eqnarray}}
\newcommand{\eeqar}{\end{eqnarray}}
\newcommand{\bit}{\begin{itemize}}
\newcommand{\eit}{\end{itemize}}
\newcommand{\benum}{\begin{enumerate}}
\newcommand{\eenum}{\end{enumerate}}
\newcommand{\barr}{\begin{array}}
\newcommand{\earr}{\end{array}}

\newcommand\eq[1]{(\ref{#1})}

\def\XXint#1#2#3{{\setbox0=\hbox{$#1{#2#3}{\int}$}
   \vcenter{\hbox{$#2#3$}}\kern-.5\wd0}}

\def\b0{\mbox{\boldmath $0$}}

\def\bk{\mbox{\boldmath $k$}}

\def\bv{\mbox{\boldmath $v$}}

\def\bC{\mbox{\boldmath $C$}}

\def\f0{\ensuremath{\mathbb{O}}}

\newcommand{\Go}{\omega}

\newcommand{\GG}{\Gamma}

\newcommand{\bmA}{\mbox{\boldmath $\mathcal{A}$}}

